\documentclass[letter,aps,prl,twocolumn,showpacs]{revtex4}
\usepackage{graphicx}
\usepackage{times}
\def\beq{\begin{equation}}
\def\eeq{\end{equation}}
\def\bey{\begin{eqnarray}}
\def\eey{\end{eqnarray}}

\def\lsim{\mathrel{\raise.3ex\hbox{$<$\kern-.75em\lower1ex\hbox{$\sim$}}}}
\def\gsim{\mathrel{\raise.3ex\hbox{$>$\kern-.75em\lower1ex\hbox{$\sim$}}}}

\begin{document}

\title{Implications of Direct Dark Matter Searches for MSSM Higgs Searches at the Tevatron}
\author{Marcela Carena, Dan Hooper and Peter Skands}
\affiliation{Fermi National Accelerator Laboratory, Batavia, IL  60510-0500}

\date{\today}

\begin{abstract}
Searches for the Minimal Supersymmetric Standard Model (MSSM) Higgs
bosons are among the most promising channels for exploring new
physics at the Tevatron. In particular, interesting regions of
large $\tan\beta$ and small $m_A$ are probed by searches for heavy
neutral Higgs bosons, $A$ and $H$, when they decay to $\tau^+\tau^-$ 
and $b\bar{b}$.  At the same time, direct searches for dark matter,
such as CDMS, attempt to observe neutralino dark matter particles
scattering elastically off nuclei. This can occur through
$t$-channel Higgs exchange, which has a large cross section in the case
of large $\tan \beta$ and small $m_A$. As a result, there is a natural
interplay between the heavy, neutral Higgs searches at the Tevatron
and the region of parameter space explored by CDMS. 
We show that if the lightest neutralino makes up the dark matter of our
universe, current limits from CDMS strongly constrain the prospects of heavy,
neutral MSSM Higgs discovery  at the Tevatron (at 3$\sigma$ with 4
fb$^{-1}$ per experiment) unless $|\mu| \gsim$ 400 GeV. The limits of
CDMS projected for 2007 will increase this constraint to
$|\mu| \gsim$ 800 GeV. On the other hand, if CDMS does
observe neutralino dark matter in the near future, it will make the
discovery of heavy, neutral MSSM Higgs bosons far more likely at the
Tevatron. 
\end{abstract}
\pacs{14.80.Cp;14.80.Ly;95.35.+d; FERMILAB-PUB-06-032-A}
\maketitle

{\it Introduction} --- Searches for Higgs bosons and supersymmetric particles are among the 
most exciting challenges for the Tevatron collider experiments. 
Of particular interest are 
searches for CP-even and CP-odd Higgs bosons with enhanced couplings 
to $b$-quarks and $\tau$-leptons. 

Tevatron limits on the production of heavy, neutral MSSM Higgs bosons have recently been published~\cite{tevcurrent}. 
The inclusive process $p\bar{p} \rightarrow A/H \, X \rightarrow
\tau^+ \tau^- \, X$ relies on $\tau$-lepton reconstruction to suppress
backgrounds, and includes both the gluon-fusion process and the 
radiation of the Higgs from a final-state $b$-quark.  The
process $p\bar{p} \rightarrow A/H  \,\, b \bar{b}$ followed
by $A/H \rightarrow b\bar{b}$ relies on the tagging of multiple
$b$-jets to isolate a signal.  The constraints from both of these
Higgs searches exclude values of $\tan\beta$ as low as $\sim 40$ for
$m_A\sim$100 GeV. While this represents only a corner
of parameter space, a wider range of models will
be tested as the integrated luminosity grows. With 4 fb$^{-1}$, values
of $\tan\beta$ as small as $\sim$30 and values of $m_A$ as large as
$\sim$250 GeV will be within reach of the Tevatron~\cite{tevfuture}. 
Tevatron searches for the decay $B_s \rightarrow \mu^+ \mu^-$ also cover a significant portion of the
$\tan\beta$-$m_A$ plane~\cite{bsmumu}.

Concurrently, dark matter experiments are also searching for supersymmetry, 
in the form of a stable neutralino. These experiments~\cite{cdms,edelweisszeplin} have begun to constrain supersymmetric models by providing limits on 
the spin-independent elastic scattering cross section of the lightest 
neutralino with nuclei. When the elastic scattering cross 
section, $\sigma_{\mathrm{SI}}$, is large enough to be detected by current 
experiments, this process is generally dominated by the $t$-channel exchange of 
CP-even Higgs bosons, $H$ and $h$, coupling to strange quarks and to 
gluons through a bottom quark loop. The cross section from this
contribution is enhanced at large $\tan\beta$ through the $s$ and $b$
Yukawa couplings. Within the MSSM at large $\tan\beta$, the mass of
the CP-odd Higgs is related to that of the CP-even Higgs with enhanced
couplings to down-type fermions, hence this cross section is also
enhanced for small values of $m_A$.

Comparing direct dark matter experiments to Tevatron searches, we see
that the prospects for both of these techniques depend to a certain extent
on the same parameters, $\tan\beta$ and~$m_A$. 
We explore the interplay of direct dark matter searches and the Tevatron Higgs 
searches and find that, modulo certain 
caveats discussed below, current and future constraints from CDMS limit the prospects for the discovery 
of neutral Higgs bosons with enhanced couplings to down-type fermions
at the Tevatron. A positive detection in the near future by CDMS, on
the other hand,  would be very encouraging for Tevatron searches.

{\it Neutralino Elastic Scattering Cross Section} --- The CDMS experiment is primarily sensitive to the neutralino's spin-independent (scalar) 
elastic scattering cross section:
\begin{equation}
\label{sig}
\sigma_{SI} \approx \frac{4 m^2_{r}}{\pi} [Z f_p + (A-Z) f_n]^2,
\end{equation}
where $m_r=m_N m_{\chi}/(m_N+m_{\chi})$ is the reduced mass, $Z$ is the 
atomic number of the nucleus, $A$ is the atomic mass of the nucleus, 
and $f_p$ and $f_n$ are the neutralino couplings to protons and neutrons, 
given by:
\begin{equation}
f_{p,n}=\sum_{q=u,d,s} f^{(p,n)}_{T_q} a_q \frac{m_{p,n}}{m_q} + \frac{2}{27} f^{(p,n)}_{TG} \sum_{q=c,b,t} a_q  \frac{m_{p,n}}{m_q},
\label{feqn}
\end{equation}
where $a_q$ are the neutralino-quark couplings and 
$f^{(p)}_{T_u} \approx 0.020\pm0.004$,  
$f^{(p)}_{T_d} \approx 0.026\pm0.005$,  
$f^{(p)}_{T_s} \approx 0.118\pm0.062$,  
$f^{(n)}_{T_u} \approx 0.014\pm0.003$,  
$f^{(n)}_{T_d} \approx 0.036\pm0.008$, and 
$f^{(n)}_{T_s} \approx 0.118\pm0.062$ \cite{scatteraq}. The first term in Eq.~\ref{feqn} 
corresponds to interactions with the quarks in the target, either through $t$-channel 
CP-even Higgs exchange, or $s$-channel squark exchange. The second
term corresponds to interactions with the gluons in the target through a quark/squark 
loop diagram. $f^{(p)}_{TG}$ is given by $1 -f^{(p)}_{T_u}-f^{(p)}_{T_d}-f^{(p)}_{T_s} 
\approx 0.84$, and analogously, $f^{(n)}_{TG} \approx
0.83$. 

The contribution to the neutralino-quark coupling from Higgs exchange is 
given by~\cite{scatteraq}:
\begin{eqnarray}
&~&a_q^{(\rm{Higgs})}  =  \nonumber \\
&-& \frac{g_2 m_{q}}{4 m_{W} B} \left[ Re \left( 
\delta_{1} [g_2 N_{12} - g_1 N_{11}] \right) D C \left(\frac{1}{m^{2}_{h}} - \frac{1}{m^{2}_{H}} \right) \right. \nonumber \\
&+&  Re \left. \left( \delta_{2} [g_2 N_{12} - g_1 N_{11}] \right) \left( 
\frac{D^{2}}{m^{2}_{h}}+ \frac{C^{2}}{m^{2}_{H}} 
\right) \right].
\label{aq}
\end{eqnarray}
For up-type quarks, $\delta_{1} = N_{13}$, $\delta_{2} = N_{14}$, 
$B = \sin{\beta}$, $C = \sin{\alpha}$ and $D = \cos{\alpha}$, 
whereas for down-type quarks, $\delta_{1} = N_{14}$, $\delta_{2} = -N_{13}$, 
$B = \cos{\beta}$, $C = \cos{\alpha}$ and $D = -\sin{\alpha}$. 
$\alpha$ is the Higgs mixing angle. $N^2_{11}$, $N^2_{12}$, $N^2_{13}$ and 
$N^2_{14}$ are the bino, wino and two Higgsino fractions of the lightest 
neutralino, respectively.

Supersymmetric models which are within the current or near-future
reach of CDMS generally have an elastic scattering cross section that
is dominated by CP-even Higgs exchange. For illustration, consider a bino-like
neutralino (with a small Higgsino admixture) and large to moderate
$\tan\beta$ and $\cos\alpha \sim 1$. In this case, the neutralino-nucleon cross
section from CP-even Higgs exchange is approximately given by: 
\begin{eqnarray}
\sigma_{SI}&\sim& \frac{0.1 \,m^4_p\, g^2_1\, g^2_2\, N^2_{11}\, N^2_{13}\, \tan^2\beta}{4 \pi\, m^2_W\, m^4_A} \nonumber \\
\sim 4&\times& 10^{-7} \, \rm{pb} \bigg(\frac{N^2_{11}}{0.9}\bigg)
\bigg(\frac{N^2_{13}}{0.1}\bigg) \bigg(\frac{300 \,
  \rm{GeV}}{m_A}\bigg)^4  \bigg(\frac{\tan\beta}{50}\bigg)^2~,
\label{naive}
\end{eqnarray}
where the mass of the CP-even Higgs with enhanced couplings to
down-type fermions is approximately $m_A$. 
Below, we shall express the neutralino mass and composition in terms of 
$M_1=M_2/2$ and $\mu$. Generally, the relation is not compact, but {\em e.g.} in
the limiting case of a light, bino-like LSP ($N^2_{11}
\approx 1$), one has  $N^2_{13} \approx \sin^2 \theta_W \sin^2 \beta \,
m^2_Z/\mu^2$. 

Eq.~\ref{naive} demonstrates that if 
$m_A$ and $\tan \beta$ are within the range of Tevatron
searches, then a substantial elastic scattering cross section can
be expected for the lightest neutralino, unless it is a very
pure bino ({\it ie.} $|\mu|$ is very large).

\begin{figure}

\resizebox{8.4cm}{!}{\includegraphics{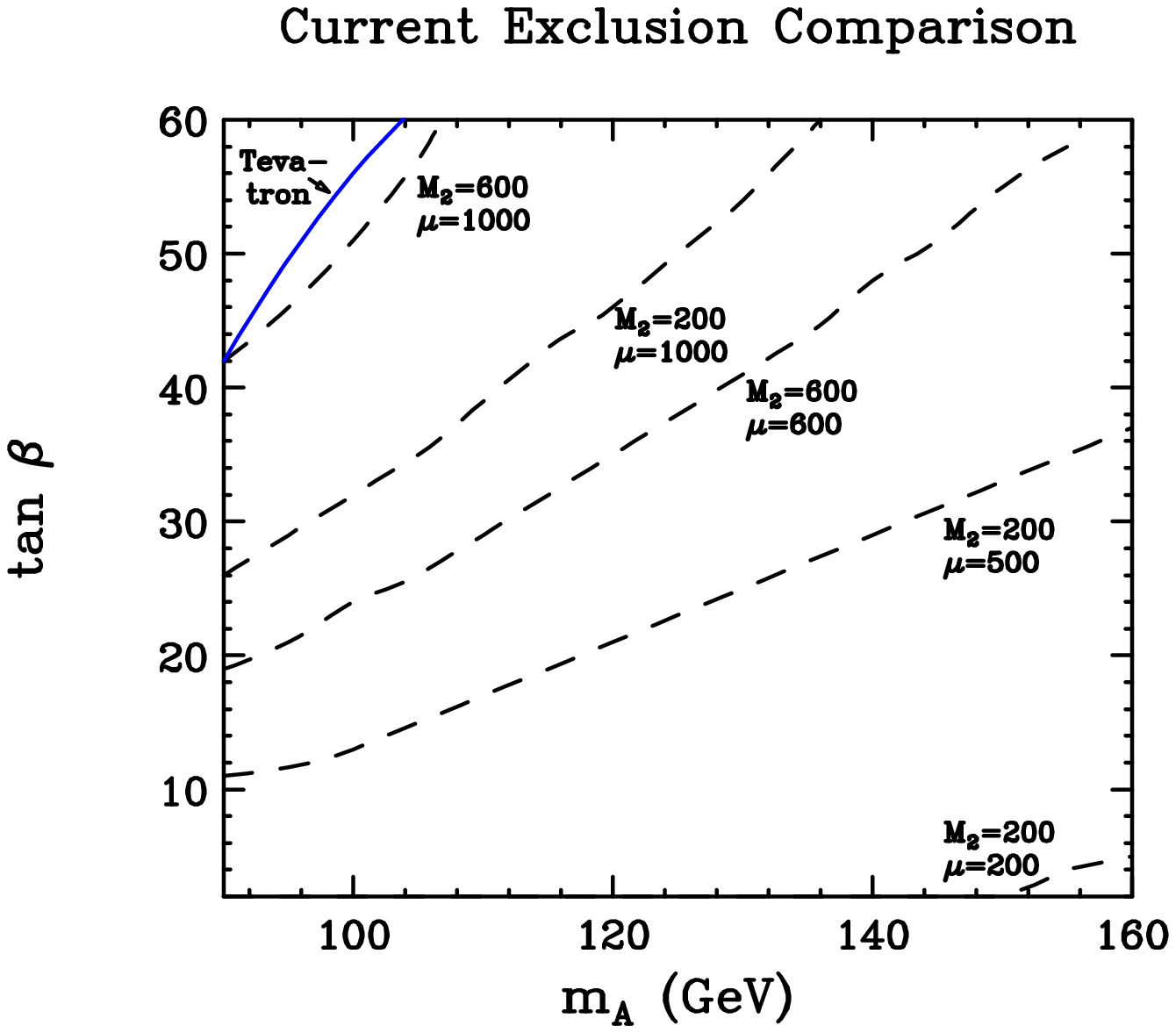}} \\
\vspace{0.5cm}
\resizebox{8.4cm}{!}{\includegraphics{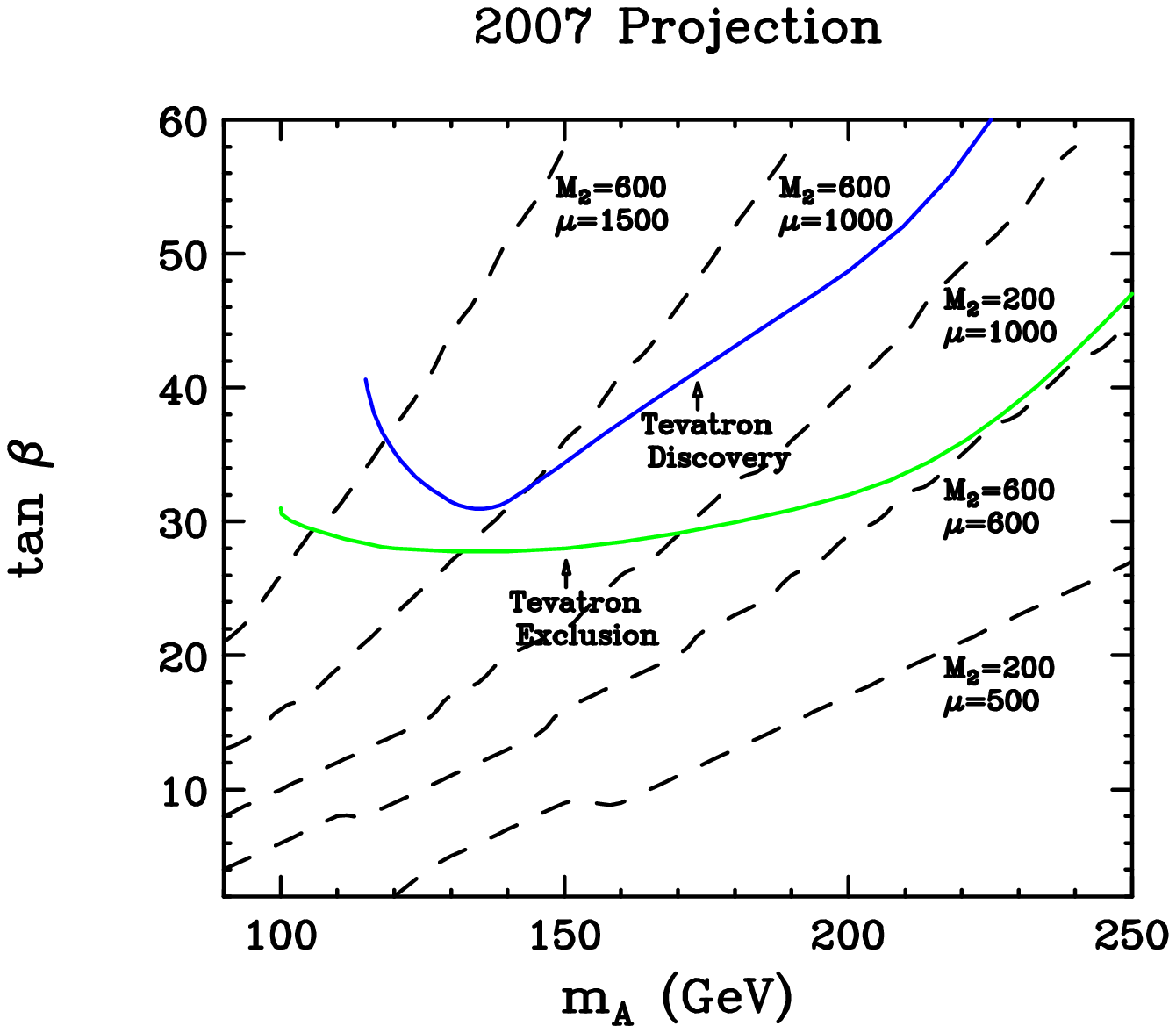}}

\caption{Top-frame: The current Tevatron limit from 
$p\bar{p}\rightarrow A/H \,\, X \rightarrow \tau^+ \tau^- \,\, X$ 
\cite{tevcurrent} 
compared to the currently excluded 
regions from CDMS for various combinations of $M_2$ and $\mu$. 
Bottom-frame: The projected Tevatron 3$\sigma$ discovery
and  95\% exclusion reach for 
$p\bar{p}\rightarrow A/H \,\,X \rightarrow \tau^+ \tau^- \,\,X$
\cite{tevfuture} 
compared to the 2007 projection of 
the CDMS limits.  
\label{curpro}}
\end{figure}

{\it Implications of CDMS} --- CDMS currently provides the strongest constraints on $\sigma_{SI}$. For a 50--100 GeV neutralino, this result
excludes $\sigma_{SI} \gsim 2 \times 10^{-7}$ pb, whereas the limit
is about a factor of ten weaker for a 1 TeV neutralino \cite{cdms}.  

In the top frame of Fig.~\ref{curpro}, we have plotted as a 
solid line the current exclusion limit of the Tevatron for 
$p\bar{p}\rightarrow A/H \, X \rightarrow \tau^+ \tau^-\, X$ in the 
$\tan\beta-m_A$ plane and compared this to the current limits 
from CDMS, for various choices of $M_2$ and $\mu$. The Tevatron
constraint from the inclusive $\tau^+\tau^-$ channel is quite robust
against variations of the MSSM parameters, while the channel
$p\bar{p}\rightarrow A/H \,\, b\bar{b}$ followed by 
$A/H \rightarrow b\bar{b}$ is more susceptible to radiative
corrections and is weaker, unless both $|\mu|$ is large 
and $\mu M_3 < 0$~\cite{rad}.

\begin{figure}

\resizebox{8.4cm}{!}{\includegraphics{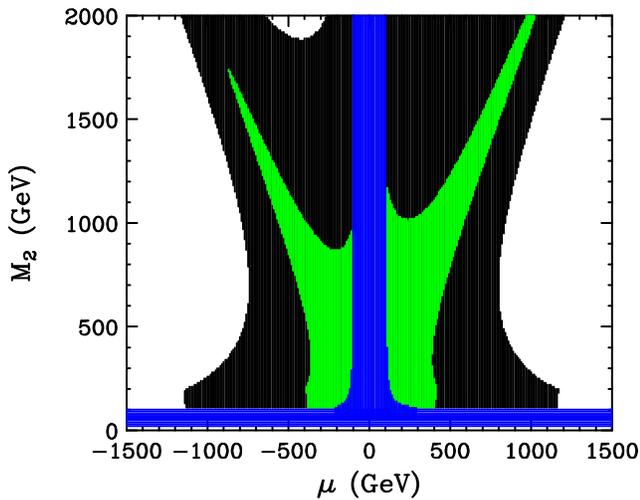}} \\
\caption{The regions in the $M_2$-$\mu$ plane in which the possibility 
of discovering heavy, neutral MSSM Higgs boson at the Tevatron 
(4 fb$^{-1}$ per experiment) through $p\bar{p} \rightarrow A/H \, X
\rightarrow \tau^+ \tau^- \, X$ is excluded due to current CDMS limits 
(light shaded) and the projected 2007 CDMS limits (black). 
The shaded region along the bottom of the figure and extending 
upward for small $\mu$ is excluded by LEP chargino limits~\cite{LEPcharginolimits}.}
\label{scan}
\end{figure}

$\sigma_{\rm{SI}}$ was calculated using the 
DarkSUSY program \cite{darksusy} assuming the central values 
of the $f_T$'s appearing in Eq.\ref{feqn}. The squarks have been 
decoupled and the GUT-relation between the gaugino masses 
was adopted. Note also that we do not address the neutralino relic
abundance in Figs.~\ref{curpro} and~\ref{scan}, since we have only
specified those supersymmetric parameters which are relevant to
elastic scattering through Higgs exchange.

In the lower frame of Fig.~\ref{curpro}, we show the projected 
3$\sigma$ discovery reach at the Tevatron (4 fb$^{-1}$ per 
experiment) compared with the 2007 projected limits from 
CDMS~\cite{cdmsfuture}. For a wide range of $M_2$ and $\mu$, 
we find that CDMS is able to test the entire region of the 
$\tan\beta-m_A$ plane in which the Tevatron will be capable of 
observing $p\bar{p}\rightarrow A/H \, X \rightarrow \tau^+ \tau^- \, X$.

It is clear that (unless $\mu \gg M_2$) the lack of a signal at CDMS disfavors the possibility of discovering heavy, 
neutral MSSM Higgs bosons at the Tevatron. We show these results in the $M_2$-$\mu$ plane in Fig.~\ref{scan}. For the 
models in the lightly shaded regions, $A/H$ is not expected to be
observed in the inclusive $\tau^+\tau^-$ channel, for any values of 
$\tan\beta$ and $m_A$, given the current constraints 
from CDMS. This would be expanded to the black regions, if indeed CDMS observes no signal in~2007.  

If $\chi^0$ had a large Higgsino component, such as in the upper
region of Fig.~\ref{scan} where $M_2/2 \gg |\mu|$, then it would be
produced below the measured dark matter density~\cite{wmap}.  
Consequently, we focus on the lower region of the plot, in which
$M_2/2 \lsim |\mu|$.  In the case when
$M_2/2 \ll |\mu|$, the lightest neutralino is mostly bino-like, 
and the elastic scattering cross section is 
gradually reduced. For example, for $M_2=$200 GeV, the Higgsino fraction 
of the lightest neutralino is approximately 15\%, 1\% and 0.2\% for 
$\mu=$200, 500 and 1000 GeV, respectively, corresponding to $\sigma_{\rm{SI}}
\sim 6 \times 10^{-7}$, $4 \times
10^{-8}$ and $10^{-8}$ pb, for $\tan \beta=50$ and $m_A=300$ GeV (see Eq.~\ref{naive}).

A positive signal at CDMS would be very 
encouraging for Tevatron Higgs searches. For example, if a neutralino 
were detected at CDMS with $\sigma_{SI}\sim 10^{-7}$ pb,  
then 3$\sigma$ evidence for $A/H$ in the inclusive $\tau^+\tau^-$ channel
would be obtained as long as the Higgsino fraction of the lightest neutralino 
is greater than about $0.5\%$ and $m_A$ is heavier than about 140 GeV (as inferred from Fig.~\ref{curpro} and Eq.~\ref{naive}). On the other hand, evidence for the production of heavy neutral Higgs 
bosons at the Tevatron, without an observation
at CDMS by~2007, could give very valuable information about the MSSM particle
spectrum.  In particular, it would suggest that $|\mu|$ is large, {\em e.g.} 
greater than about~800~GeV.

{\it Caveats} --- These conclusions are subject to a number of 
assumptions. Most obviously, if the dominant component of our 
universe's dark matter is not made up of neutralinos, then the 
constraints placed by CDMS do not affect collider searches.

The results from CDMS involve substantial
astrophysical uncertainties, including the
local dark matter density, which we have taken to be
$0.3$~GeV/cm$^3$, as implied by the dynamics of our galaxy. Halo
profiles consistent with observations have been proposed in which the
dark matter density at the radius of the solar circle is as large as
0.8~GeV/cm$^3$ and as small as 0.2~GeV/cm$^3$. Experiments such as
CDMS measure the product of the local dark matter density and the 
cross section averaged over the relative neutralino velocity.

If dark matter is not distributed homogeneously, but in dense
clumps or tidal streams, then the density and velocity distribution of
neutralinos at Earth could be different from the values we have
used. Simulations suggest, however, that the dark matter distribution
in the local vicinity consists of a superposition of a very large
number of substructures, making substantial deviations from homogeneity unlikely~\cite{sim}.  

\begin{figure}
\resizebox{8.4cm}{!}{\includegraphics{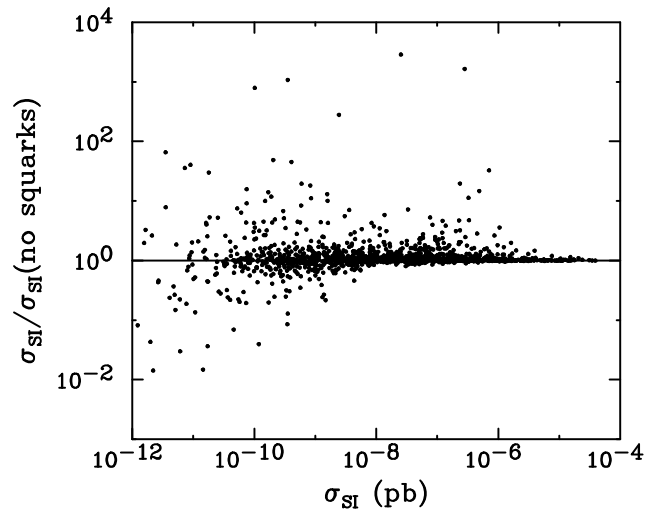}} \\
\caption{The neutralino's elastic scattering cross section compared to 
the value found if all squark contributions are neglected, for a random 
sample of supersymmetric models. Each point shown respects all current 
collider constraints and does not predict a thermal density of neutralinos 
in excess of WMAP measurements.}
\label{scatter}
\end{figure}

We have neglected squark exchange diagrams in 
the calculation of $\sigma_{rm{SI}}$ thus far. 
In Fig.~\ref{scatter} we 
demonstrate that these contributions, once included, 
generally increase rather than 
decrease the cross section, thus making our conclusions stronger.
The figure shows the elastic scattering cross section for a random 
sample of lightest neutralinos compared to the value found if all 
squark contributions are neglected. Here we have scanned over $M_2$, 
$\mu$, $\tan\beta$, $m_A$, sfermion masses and trilinear couplings 
(up to 2 TeV). $M_1$ and $M_3$ have been set by the GUT 
relationship. In the range of cross sections for which CDMS is likely to be sensitive to in the near future 
($\sigma_{SI} \gsim 10^{-8}$ pb), neglecting squarks either has little
effect or slightly underestimates the cross section in each model
found. 

We also consider our MSSM model assumptions, and alternatives 
which might lead to smaller elastic cross sections.   For example, 
we assumed the GUT relationship between the gaugino masses  
($M_1:M_2:M_3 \approx 1:2:7.5$), which leads to a very small wino
component for the lightest neutralino. If, instead, we consider
$M_2 \lsim M_1$, then the lightest 
neutralino could be wino-like (such as in models of 
Anomaly Mediated Supersymmetry Breaking), and the cross section
will be modified accordingly~\cite{HooperWang}. We have also assumed all MSSM parameters to be purely real.  If, for
example, $\mu$ has a non-trivial phase, then the Higgs states will
mix and the Higgs-neutralino couplings can vary widely with that 
phase~\cite{cpviolation}.  In such a scenario, the elastic scattering 
of neutralinos could be substantially suppressed.

Also, throughout our analysis we have
used the central values \cite{scatteraq} for the $f^{(p,n)}_T$ parameters appearing 
in Eq.~\ref{feqn}. For $\sigma_{SI}$, the dominant contributions come
from $f_{T_s}$ and $f_{TG}$. Since these are related by a sum rule, 
our results are slightly more stable than would be inferred from the
large uncertainty on $f_{T_s}$ alone. Varying $f_{T_s}$ by
$1\sigma$, for example, causes a factor of roughly 2 change in
$\sigma_{SI}$, in either direction. 
Note that a larger $\pi$-nucleon
$\Sigma$ term \cite{newsigma} would increase $f_{T_s}$ and thereby
$\sigma_{SI}$, hence the values used here are
conservative in this regard.

Finally, we point out that we have neglected radiative corrections 
to the down-type Yukawa couplings, which can be important at the largest 
values of $\tan \beta$ considered here. They will impact slightly
the discovery reach of the Tevatron Higgs searches in the inclusive-$\tau$
channel, and more significantly the value of~$\sigma_{\mathrm{SI}}$~\cite{rad},
although the impact is generally small compared to the astrophysical 
uncertainties discussed above.

{\it Summary and Conclusion} --- In this letter, we have explored the interplay of direct dark matter 
searches such as CDMS and the search for heavy, neutral MSSM Higgs bosons at 
the Tevatron. Both search techniques are most sensitive to 
supersymmetric models with large $\tan\beta$ and small $m_A$. 
We find that, modulo the caveats we have discussed, for small and moderate values of $|\mu|$, this region of
MSSM parameter space is being probed first by CDMS. 
If a neutralino signal is seen in the near future by CDMS, then the expectations for the discovery of heavy, neutral Higgs bosons at the Tevatron
will be high.  On the other hand, the lack of a signal in CDMS
would suggest that the observation of Higgs bosons will be unlikely,
unless the lightest neutralino is a nearly pure bino, as would be
the case if $|\mu| \gg M_2/2$. The current constraints from 
CDMS disfavor the discovery of heavy, neutral MSSM Higgs bosons at the 
Tevatron unless $|\mu|\gsim$ 400 GeV, and the 2007 projected limits of 
CDMS will extend this to $|\mu|\gsim$ 800 GeV. 

Collider searches for heavy, neutral MSSM Higgs bosons and direct dark
matter searches can, together, be used to extract valuable information
about the supersymmetric spectrum. An observation of
$p\bar{p}\rightarrow A/H \rightarrow \tau^+ \tau^-$ at the Tevatron
along with an accompanying discovery at CDMS, for example, could
potentially be used to infer the higgsino fraction of the lightest
neutralino. An observation of $p\bar{p}\rightarrow A/H \rightarrow
\tau^+ \tau^-$ at the Tevatron without an accompanying discovery at
CDMS, on the other hand, would imply a very small higgsino fraction
for the lightest neutralino (a large value of $|\mu|$), or that
the nature of dark matter is not as assumed.


DH is supported by the US Department of Energy and by NASA grant NAG5-10842. MC and PS are supported by the US Department of Energy grant DE-AC02-76CHO3000.

\end{document}